\documentclass[a4paper]{jpconf}
\usepackage{graphicx}
\begin{document}
\title{DEAP-3600 Dark Matter Search at SNOLAB}
\author{M. G. Boulay for the DEAP Collaboration}
\address{Queen's University, Department of Physics, Engineering Physics and Astronomy, Kingston, Ontario, Canada, K7L 3N6}
\ead{mark.boulay@queensu.ca}

\begin{abstract}
The DEAP-3600 detector, currently under construction at SNOLAB, has been designed to achieve extremely low background rates from all sources, including $^{39}$Ar $\beta$-decays, neutron scatters (from internal and external sources), surface $\alpha$-contamination and radon. An overview of the detector and its sensitivity are presented.
\end{abstract}

\begin{figure}
\hspace{1in}\includegraphics[width=5.5in]{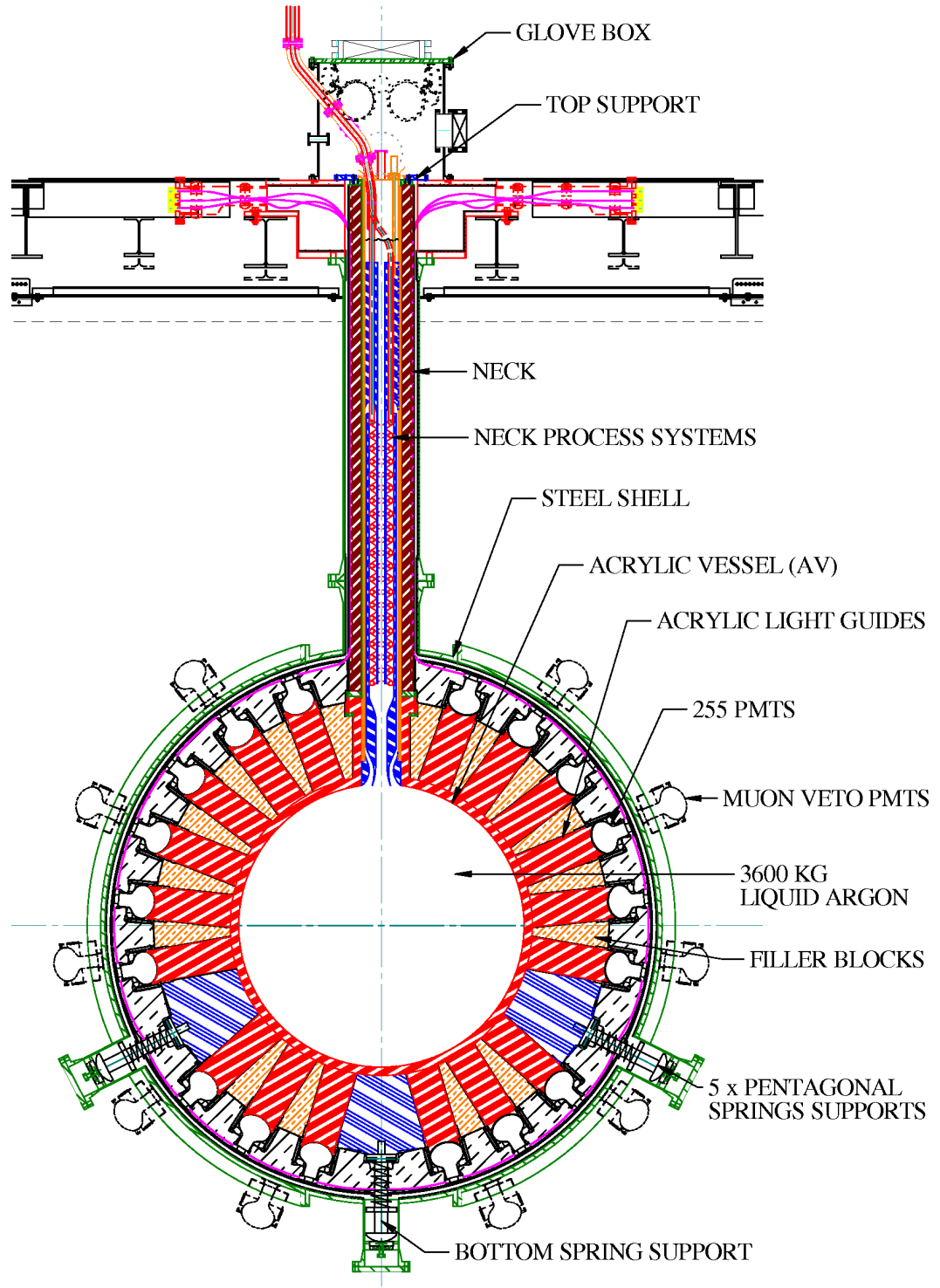}
\caption{The DEAP-3600 detector. The acrylic vessel has in inner radius of 85 cm and holds 3600 kg of liquid argon, which is viewed by 255 8-inch high quantum efficiency photomultiplier tubes (PMTs) through 50-cm lightguides. The lightguides provide neutron shielding and thermal insulation between the cryogenic acrylic vessel and ``warm'' PMTs.  The inner detector is housed in a large stainless steel spherical vessel, which itself is immersed in an 8 meter diameter water shielding tank.}
\label{deap_detector}
\end{figure}

DEAP-3600 will perform a dark matter particle search on argon with sensitivity to the spin-independent WIMP-nucleon cross-section of 10$^{-46}$ cm$^{2}$, a factor of approximately one hundred increase over current experiments.  DEAP has been under development since 2005 and is a collaborative effort including over 60 researchers. The detector, shown in Fig.~\ref{deap_detector} is conceptually simple:  a large spherical volume of liquid argon contained in a transparent acrylic vessel is viewed by 255 photomultiplier tubes that detect scintillation light pulses generated in the argon.   DEAP-3600 will allow a three-year exposure of a one-thousand kilogram (fiducial) target of liquid argon in the SNOLAB Cube Hall, which itself has extremely suppressed backgrounds due its 6000 m.w.e. depth. Extreme care has gone into selection of detector materials, including their exposure to radon either during fabrication, or installation at SNOLAB, to ensure background target levels will be met.  The acrylic vessel and lightguide acrylic have undergone extensive quality control to ensure radiopurity of the bulk acrylic; of particular concern was the diffusion of radon into the acrylic monomer leading to a buildup of long-lived $^{210}$Pb, which would lead to a source of ($\alpha$,n) neutron backgrounds in the inner detector.  After the acrylic vessel has been fabricated, it will be sealed from the mine environment, and while purged with low-radon gas, a several-micron layer will be removed with a custom resurfacer to ensure the detector surface is as radiopure as the bulk acrylic.  A custom large-area vacuum-deposition source developed by the DEAP collaboration will then be used to deposit tetraphenyl butadiene (TPB) wavelength shifter uniformly over the approximately 10 m$^2$ acrylic detector surface, before the vessel is cooled and filled with liquid argon.  The detector design parameters and background targets are shown in Table~\ref{table1}.

\begin{figure}[h]
\hspace{-0.2in}\includegraphics[width=3.65in]{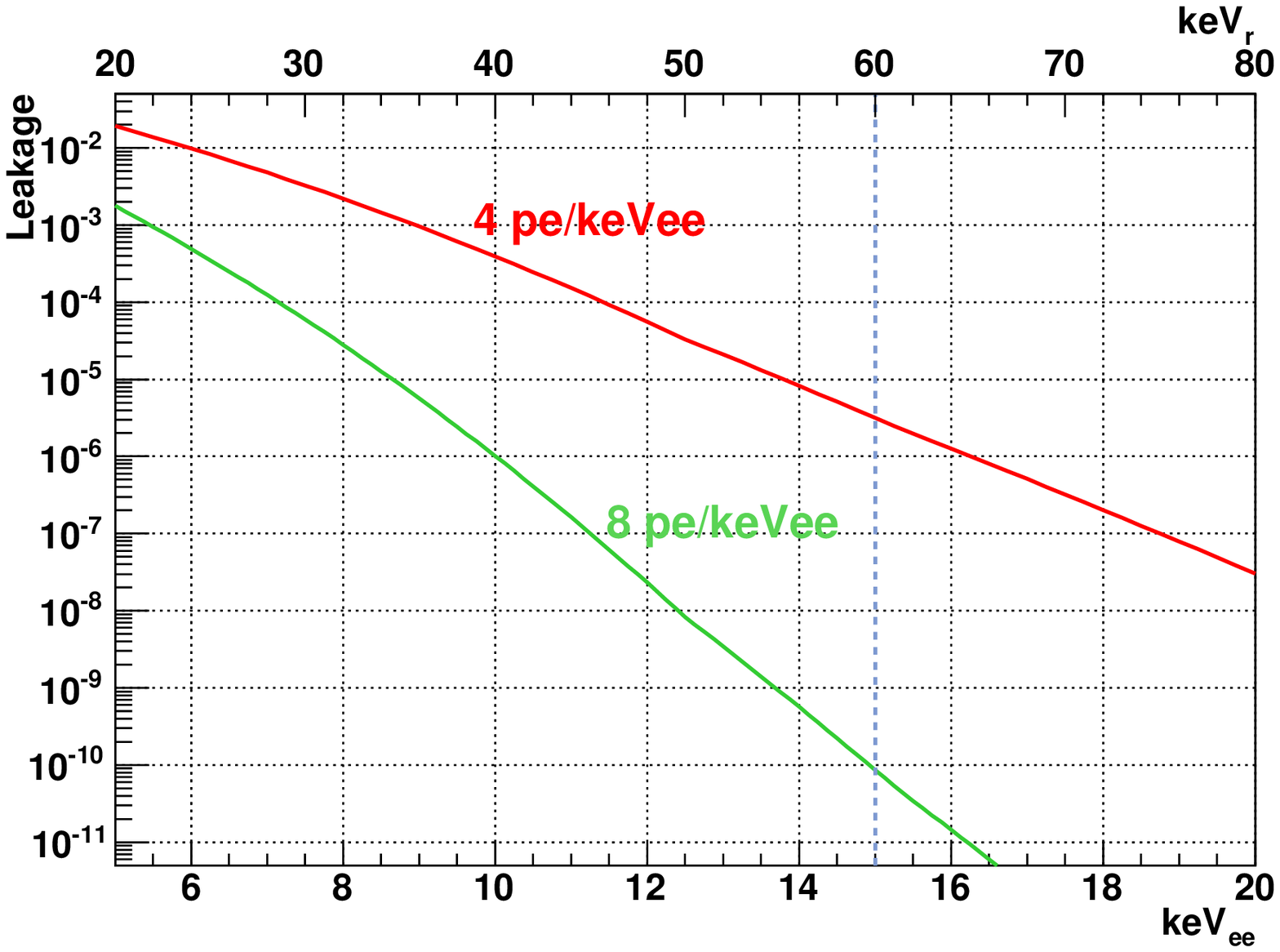}
\hspace{-0.3in}\includegraphics[width=3.65in]{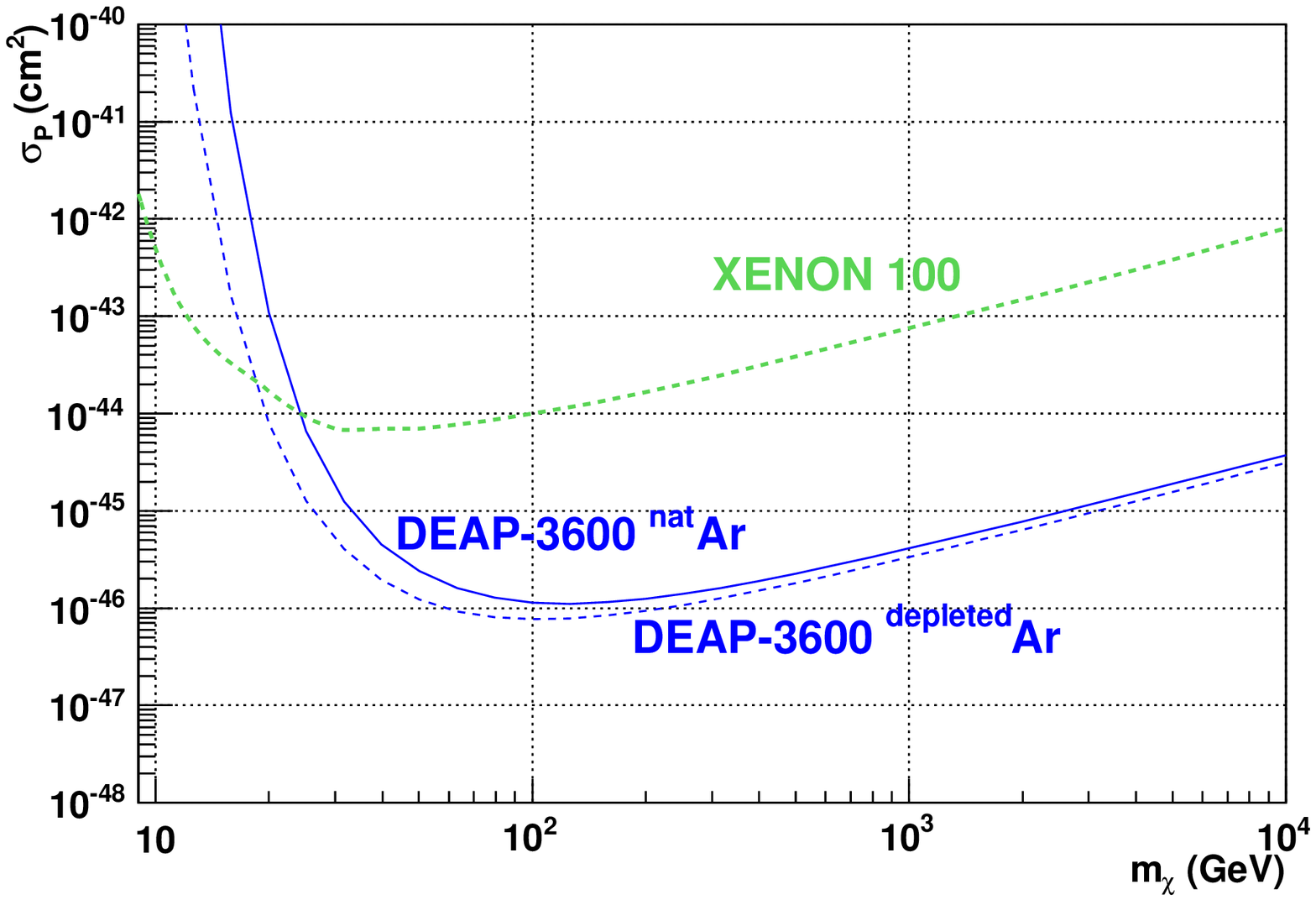}
\caption{Pulse-shape discrimination and sensitivity with liquid argon. (Left) Discrimination of $\beta$ events versus energy for light yields of 4 and 8 photoelectrons per keV$_{\rm{ee}}$.  The higher light yield improves discrimination by approximately $5 \times 10^{4}$ at the 15 keV$_{\rm ee}$ threshold. (Right) 90\% exclusion limit for DEAP-3600 with natural and depleted argon.  Also shown for reference is the current XENON-100 exclusion limit~\cite{xenon100}.  }
\label{deap_psd}
\end{figure}

\begin{figure}[h]
\hspace{-0.2in}\includegraphics[width=3.65in]{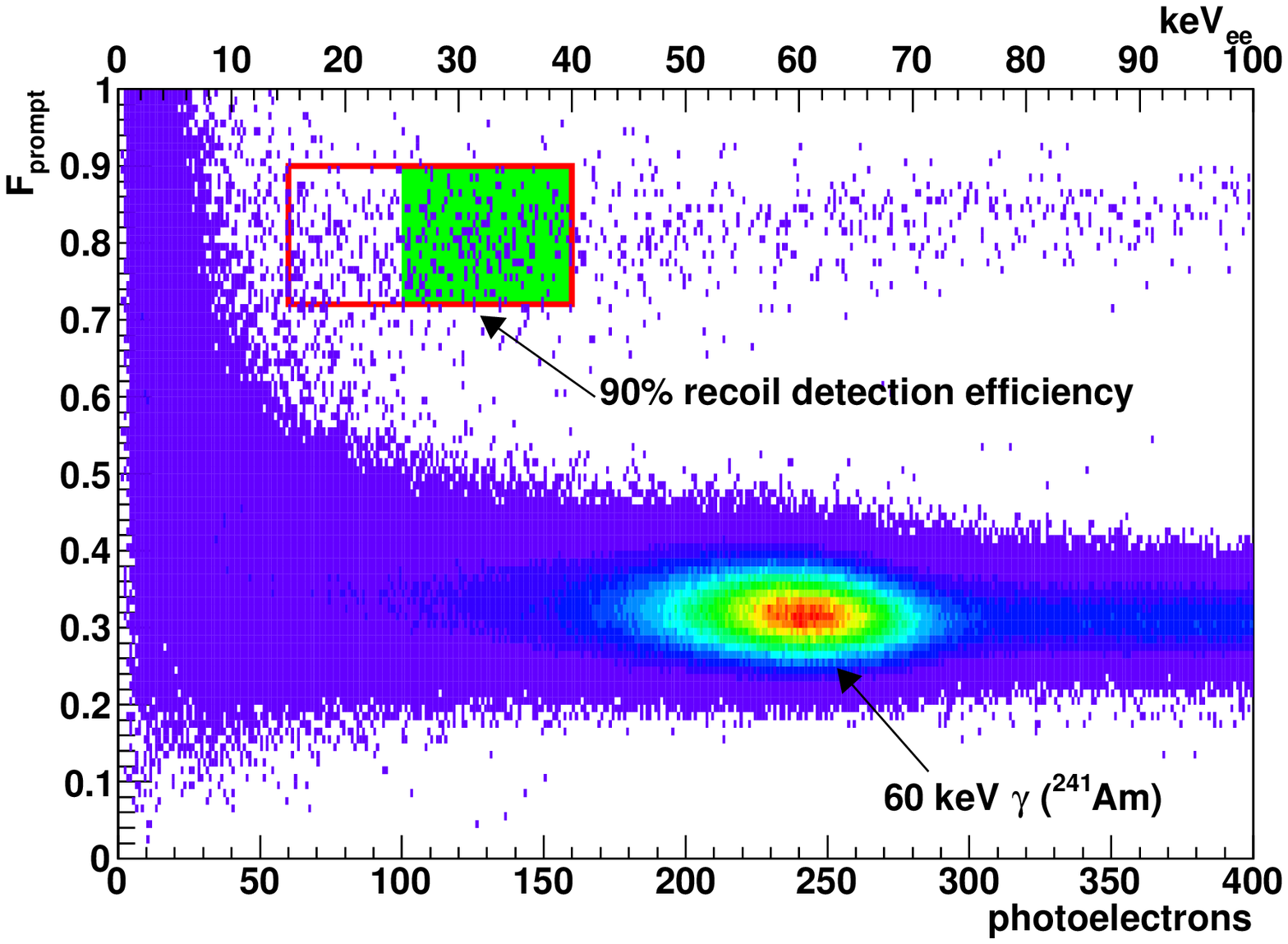}
\hspace{-0.3in}\includegraphics[width=3.65in]{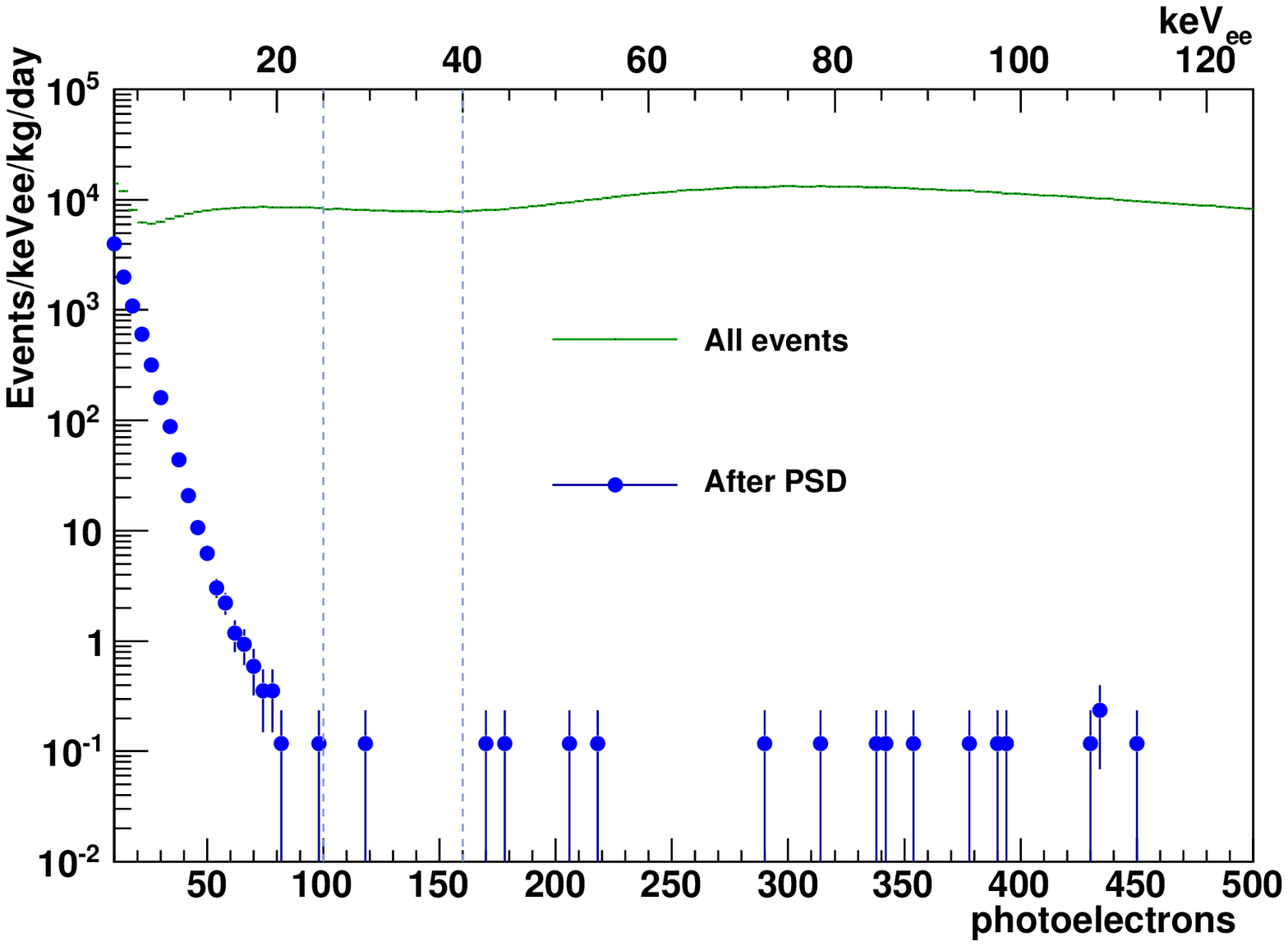}
\caption{Results from the DEAP-1 7-kg liquid argon prototype detector at SNOLAB. (Left) Neutrons (high-F$_{\rm{prompt}}$ band) and $\gamma$'s (low-F$_{\rm{prompt}}$ band) from an untagged Am-Be source. The boxes show F$_{\rm{prompt}}$ regions with 90\% detection efficiency, with DEAP-3600 and DEAP-1 thresholds. (Right) Backgrounds in the DEAP-1 prototype.  The low-energy ``wall'' is from $\gamma$ backgrounds that are not removed by PSD; high-energy events are from radon and surface contamination.  In the energy region from 25 to 40 keV$_{\rm{ee}}$, the backgrounds correspond to approximately 100 $\mu$Bq/m$^{2}$.}
\label{ambe_run}
\end{figure}

\begin{table}
\begin{tabular}{l|l} \hline
Parameter & Value \\ \hline
Light Yield & 8 photoelectrons per keV$_{\rm{ee}}$ \\
Nuclear Quenching Factor & 0.25 \\
Analysis Threshold & 15 keV$_{\rm ee}$, 60 keV$_{\rm r}$ \\
Total Argon Mass (Radius) & 3600 kg (85 cm) \\
Fiducial Mass (Radius) & 1000 kg (55 cm) \\
Position Resolution at threshold (conservative, design spec.) & 10 cm\\
Position Resolution at threshold (ML fitter) & $<$ 6.5 cm \\ \hline
Background Specifications & Target \\ \hline
Radon in argon & $<$ 1.4 nBq/kg\\ 
Surface $\alpha$'s (tolerance using conservative pos. resolution) & $< 0.2~\mu$Bq/m$^2$ \\
Surface $\alpha$'s (tolerance using ML pos. resolution) & $< 100~\mu$Bq/m$^2$\\
Neutrons (all sources, in fiducial volume) & $<$ 2 pBq/kg \\
$\beta\gamma$ events, dominated by $^{39}$Ar (after PSD) & $<$ 2 pBq/kg \\
Total Backgrounds & $<$ 0.6 events in 3 tonne-years \\ \hline
\end{tabular}
\caption{\label{table1}Detector design parameters and background targets. The analysis threshold is set by the requirement for sufficient PSD of $^{39}$Ar. The 10-cm position resolution is from a centroid charge resolution reconstruction algorithm.  A more sophisticated maximum likelihood (ML) fitter improves the position resolution significantly, and allows an increased $\alpha$ surface background of approximately $100$ $\mu$Bq/m$^{2}$ (a level already demonstrated with DEAP-1).  The target background specifications correspond to approximately 1 event per 5 Gg-days for each source.}
\end{table}

Mitigation of the large $^{39}$Ar background, approximately 1 Bq/kg of argon~\cite{Looslipaper}, will be accomplished using pulse-shape discrimination (PSD) of the scintillation signal, as described in~\cite{boulay_astro,Lippincott:2008ad,deap1psd}.  All PMT signals will be digitized with CAEN v1720 digitizers. PSD in argon depends strongly on the number of detected photoelectrons (pe).  Fig.~\ref{deap_psd} shows the expected PSD using the technique outlined in~\cite{deap1psd}, for light yields of 4 or 8 pe/keV$_{\rm{ee}}$.  This strong dependence on light yield (a factor of approximately $5 \times 10^{4}$ difference in PSD between 4 and 8 pe/keV$_{\rm{ee}}$) drove the design decision towards a single-phase detector, in order to pursue a very high photocathode coverage and thus maximize light collection and background discrimination.  Given this PSD dependence on energy, a reduction of the $\beta/\gamma$ background by a factor of ten allows a threshold reduction of 1.25 keV$_{\rm ee}$.  Acosta-Kane et al have demonstrated the possibility of argon with suppressed $^{39}$Ar levels~\cite{GalbiatiPaper}, and we are working in collaboration with the Princeton group to obtain depleted argon for DEAP-3600.  90\% CL exclusion sensitivities are shown in Fig.~\ref{deap_psd}, calculated using the standard assumptions outlined in \cite{Lewin:1995rx} and with a galactic escape velocity of 544 km/s.  The increased sensitivity obtained by reducing the threshold using depleted argon is also shown in Fig.~\ref{deap_psd}; larger increases can be obtained for very low-threshold analyses.  

Fig.~\ref{ambe_run} shows data from the prototype DEAP-1 detector with 7 kg of liquid argon at SNOLAB. The inner detector has been prepared in a radon-reduced glovebox, with techniques similar to the surface preparation that will be used for DEAP-3600.  Background levels of approximately $100$~$\mu$Bq/m$^{2}$ have been demonstrated in DEAP-1; this is sufficiently low to achieve the target background from surface contamination in DEAP-3600 assuming the ML position reconstruction resolution shown in Table~\ref{table1}.

Most detector components have been fabricated.  Acrylic sheets for the vessel production have been cast at Reynolds Polymer (Asia) and thermoformed into spherical sections (in Colorado).  Light guides are being fabricated from acrylic (with measured attenuation lengths greater than 5 meters) cast at Spartech, Inc. Detector infrastructure and the shield tank have been installed at SNOLAB. Cooling for the experiment is provided by a distributed liquid nitrogen and cryocooler system. Assembly of cryogenic systems and inner detector components underground will be ongoing throughout 2012.  Commissioning and first data are scheduled for late 2013. 

\section*{References}
\bibliographystyle{unsrt}
\bibliography{deap}
\end{document}